\documentstyle[aps,pra,epsfig,twocolumn,floats]{revtex}

\begin{document}
\draft
\title{
Coupled cavities for enhancing the cross-phase modulation
in electromagnetically induced transparency
}
\author{T. Opatrn\'{y},$^{1,2}$
and D.--G. Welsch$^{1}$
 }
\address{
$^{1}$ Theoretisch-Physikalisches Institut,
Friedrich-Schiller-Universit\"at, Max-Wien-Platz 1, D-07743 Jena,
Germany \\
$^{2}$ Faculty of Science, Palack\'{y} University,
Svobody 26, CZ-77146 Olomouc, Czech Republic 
}
\date{\today}

\maketitle

\begin{abstract}

We propose an optical double-cavity resonator whose response  to a signal is
similar to that of an Electromagnetically Induced Transparency (EIT) medium. A
combination of such a device with a four-level EIT medium can serve for
achieving large cross-Kerr modulation of a probe field by a signal field. This
would offer the possibility of building a quantum logic gate based on photonic
qubits. We discuss the technical requirements that are necessary for realizing
a probe-photon phase shift of $\pi$ caused by a single-photon signal. The main
difficulty is the requirement of an ultra-low reflectivity beamsplitter and to
operate a sufficiently dense cool EIT medium in a cavity.

\end{abstract}




\section{Introduction}

Strong cross-Kerr phase modulation \cite{Schm96,Schm98} 
based on electromagnetically induced transparency (EIT)
\cite{Harris90} (for a review on EIT, see, e.g.,
\cite{Harris97,Maran98}) has been of increasing interest for 
quantum information processing. If a single photon can change
the phase of another photon, then one can use the effect for
constructing a controlled-NOT quantum gate. To realize 
a maximally large cross phase modulation (XPM) on a single-photon
level, both cavity \cite{Imam97} and free-medium
\cite{Harris99,Lukin00} regimes have been considered.

The advantage of the cavity regime is that a relatively strong
electromagnetic field can be built up inside a cavity and that
the photons can interact during a relatively long time. However,
the combination of the cavity properties with the    
ultra-low group velocity of light in the EIT medium
leads to an extremely narrow line of the transmitted light
\cite{Lukin98} which introduces some limits of the
applications that are possible in principle
\cite{Gran98,Imam98E,Gheri99}. Another disadvantage results
from the fact that even for the best mirrors that are presently
available absorption is of the same order of magnitude as
transmission \cite{Kuhn}. Thus even in the case of exact resonance,
only a fraction of about $\approx 25$\%
of the incident light is transmitted,
whereas a fraction of about $\approx 50$\%
is absorbed by the mirrors (see App.~\ref{Ap-cavity1}).
Obviously, devices giving rise to such huge losses
cannot serve for quantum information processing. 

In the free-medium regime, a probe pulse, which is desired to be
influenced by a signal pulse, moves in a free EIT medium and has thus
a very small group velocity \cite{Hau99}. The signal pulse
whose mid-frequency is sufficiently far from the medium
resonance moves with a velocity close to $c$. The interaction
time of the signal and probe is thus limited by the short overlap
time of the two pulses \cite{Harris99}. To overcome this problem,
it was suggested to use a second EIT medium for the signal so that
both signal and probe have the same (small) group velocities $v_g$
\cite{Lukin00}. Both pulses can then interact during a long time. The
disadvantage of such a scheme is that most of the signal-photon energy
is wasted dressing the atoms of the (second) EIT medium 
\cite{Mara99}
and only small fraction $\sim v_g /c$ of it is available for shifting
the phase of the probe.

In this paper, we suggest a regime in which the probe pulse moves
in a free EIT medium, whereas the signal field is confined to a
special double-cavity (Michelson-like) resonator whose linear
response is analogous to that of an EIT medium. Thus, a
strong signal can interact with the probe during a long time.
The cavity-cavity and the input-output couplings are accomplished
with weakly reflecting beam-splitters so that the losses at the
mirrors are avoided. Construction of such beam-splitters may be
a technical challenge, but could be achievable in principle.

The paper is organized as follows. In Section \ref{se2} the
double-cavity system is studied. The combined action
of the double-cavity system and the EIT medium are studied
in Section \ref{se3}, and the single-photon phase shift
is calculated. Finally, some concluding remarks are given in
Section \ref{se4}.


\section{Coupled cavities}
\label{se2}


\subsection{Linear response and ``Rabi'' splitting}
\label{se2.1}
 
\begin{figure}
\centerline{\epsfig{figure=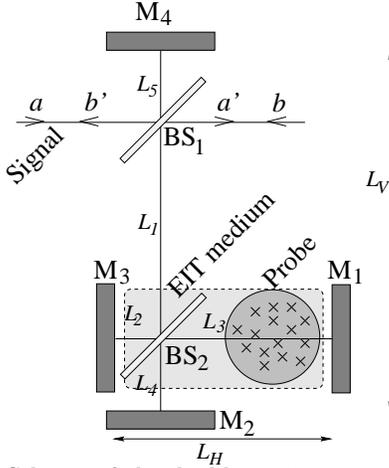,width=0.6\linewidth}}
\caption{ 
Scheme of the double-cavity system. The horizontal 
cavity is filled with an EIT medium. The signal 
enters ($a$) and leaves ($a')$ the system 
via the beam splitter BS$_1$. 
The probe propagates across the horizontal cavity in the direction
perpendicular to the plane of the plot.
}
\label{fig1}
\end{figure}

Let us consider the two-cavity scheme shown in Fig.~\ref{fig1}.
The distances are denoted by $L_1$ to $L_5$,
the lengths of the vertical and horizontal cavities being
respectively \mbox{$L_{V}$ $\!=$ $\!L_1$ $\!+$ $\!L_4$ $\!+$ $\!L_5$}
and \mbox{$L_{H}$ $\!=$ $\!L_2$ $\!+$ $\!L_3$}. The vertical
cavity is coupled to the outside field by a beam splitter BS$_1$,
and the two cavities are coupled to each other by a
beam splitter BS$_2$, with 
\begin{equation}
\label{beamsplit}
{\bf T}_l = {\hspace{1ex}t_l \hspace{1ex} ir_l \choose ir_l \hspace{1ex} t_l}
\end{equation}
($l$ $\!=$ $\!1,2$) being the  characteristic transformation
matrices of the two beam splitters. Here, $r_{l}^2$ $\!\equiv$ $\!R_{l}$
and $t_{l}^2$ $\!\equiv$ $\!T_{l}$
are respectively the reflectivities and transmissivities
satisfying the relations \mbox{$R_{l}$ $\!+$ $\!T_{l}$ $\!+$ $\!A_{l}$
$\!=$ $\!1$}, where $A_{l}$ are the absorption coefficients. 
In what follows we are interested in weakly reflecting
beam splitters, such that \mbox{$R_{l}$ $\!\ll$ $\!1$}
and \mbox{$T_{l}$ $\!\approx$ $\!1$} for
\mbox{$A_{l}$ $\!\ll$ $\!R_l$}.
The reflection and small absorption coefficients of the four
mirrors M$_{m}$ ($m$ $\!=$ $\!1,2,3,4$) are respectively $R_{Mm}$
and $A_{Mm}$, with \mbox{$R_{Mm}$ $\!+$ $\!A_{Mm}$
$\!\approx$ $\!R_{Mm}$ $\!\approx$ $\!1$}.

The input-output relations for the complex amplitudes 
$a$ ($a'$) and $b$ ($b'$) of the incoming (outgoing) waves
of wave number \mbox{$k$ $\!=$ $\!\omega /c$}
are then found by successive application of the
transformations realized by the beam splitters and the mirrors.
The result can be given in the form of
\begin{eqnarray}
 \left(  
 \begin{array}{c}
  a' \\  b'
 \end{array}
 \right) =  \left(
 \begin{array}{rr}
  G_{11} & G_{12} \\
  G_{21} & G_{22}
 \end{array}
  \right) 
  \left( 
 \begin{array}{c}
   a \\  b
 \end{array}  
 \right) , 
 \label{inout}
\end{eqnarray}
where 
\begin{eqnarray}
  G_{11} & = & t_{1}\frac{1+(1-A_1)\sqrt{1-A_{M4}}\ e^{i2kL_5} {\cal B}}
 {1+T_1 \sqrt{1-A_{M4}}\ e^{i2kL_5} {\cal B}} ,
 \label{G11}
 \\[.5ex]
 G_{12}&  = & -\frac{R_1 {\cal B}}
 {1+T_1 \sqrt{1-A_{M4}}\ e^{i2kL_5} {\cal B}} ,  
 \label{G12}
 \\[.5ex]
 G_{21} & = & \frac{R_1 \sqrt{1-A_{M4}} \ e^{i2kL_5} }
 {1+T_1 \sqrt{1-A_{M4}}\ e^{i2kL_5} {\cal B}} ,  
 \label{G21}
 \\[.5ex]
 G_{22} & =&  G_{11} ,
 \label{G22}
\end{eqnarray}
with 
\begin{eqnarray}
 {\cal B} &=& \frac{ {\cal B}_1 + {\cal B}_2 + {\cal B}_3 }{ {\cal B}_4 }, 
 \label{Bdef}
 \\[.5ex]
 {\cal B}_1 &=& (1\!-\!A_2)^2\sqrt{(1\!-\!A_{M1})(1\!-\!A_{M2})(1\!-\!A_{M3})}
 \nonumber \\[.5ex]
 && \hspace{2ex}\times\,
 e^{i2k(L_1+L_4+L_H)} ,
 \\[.5ex]
 {\cal B}_2 &=& - T_2 \sqrt{1-A_{M2}}\ e^{i2k(L_1+L_4)},
 \\[.5ex]
 {\cal B}_3 &=& R_2 \sqrt{1-A_{M3}}\ e^{i2k(L_1+L_2)},
 \\[.5ex]
 {\cal B}_4 &=& 1- \sqrt{1-A_{M1}}
 \left[ T_2 \sqrt{1-A_{M3}} \ e^{i2kL_H} \right.
 \nonumber \\
 &&\hspace{2ex} \left. - \,R_2 \sqrt{1-A_{M2}}
 \ e^{i2k(L_3+L_4)} 
 \right] .
\end{eqnarray}
Here only relations for c-numbers ($a$, $a'$, $b$, $b'$) are considered.
Note that for obtaining the operator input-output
relations, Eq.~(\ref{inout}) must be complemented with an
absorption matrix acting on some device variables\cite{Stefan}.

\begin{figure}[!t]
\noindent
\centerline{\epsfig{figure=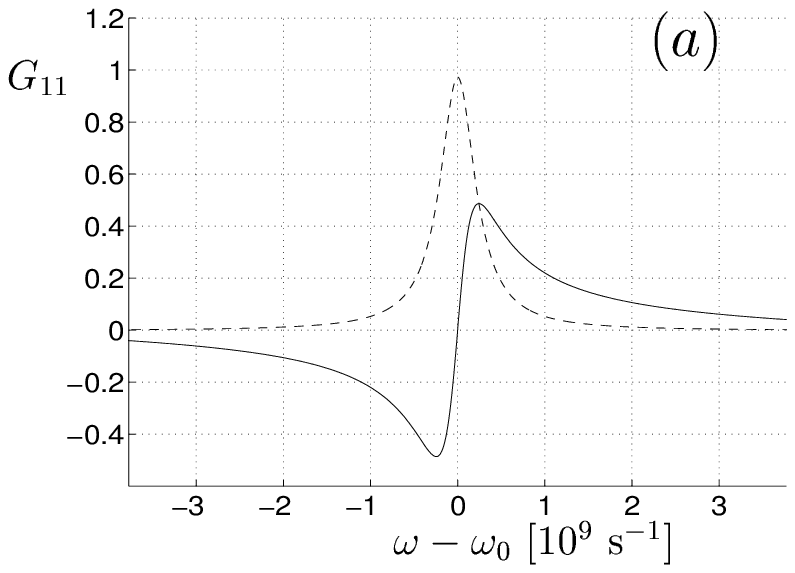,width=0.8\linewidth}}
\\[2ex]
\centerline{\epsfig{figure=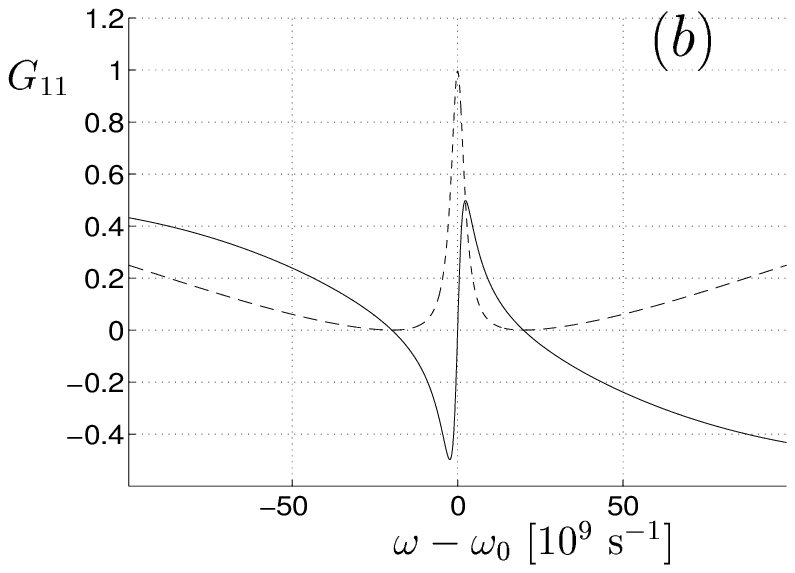,width=0.8\linewidth}}
\\[2ex]
\caption{ 
Response function $G_{11}$
for \mbox{$L_V$ $\!=$ $\!120 \lambda_0$},  
\mbox{$L_H$ $\!=$ $\!30 \lambda_0$}
(with $\lambda_0 = 2\pi /k_0 = 795\ $nm, $\omega_0 = k_0 c$),
$A_{Mm}= 10^{-6}$,
\mbox{$R_1$ $\!=$ $\!10\%$}, and \mbox{$R_2$ $\!=$ $\!10^{-6}$} (a),
\mbox{$R_2$ $\!=$ $\!10^{-5}$} (b).
Dashed line: Re$\,G_{11}$,
full line: Im$\,G_{11}$.  
}
\label{fig2}
\end{figure}
An example of the response function of the transmitted signal
$G_{11}$ is shown in Fig.~\ref{fig2}. It corresponds to the case
where the two cavities have the same resonance wave-number \mbox{$k_0$
$\!=$ $\!n_V \pi/ L_V$ $\!=$ $\!n_H \pi/L_H$}, with
\mbox{$n_V$,\,$n_H$} being integers. Were there only the
vertical
cavity (\mbox{$R_2$ $\! =$ $\! 0$}), a signal of wave
number $k=k_0$ would be reflected, and thus \mbox{$|G_{11}|$
$\! \approx$ $\! 0$} would be observed.
Due to the coupling introduced by the beam splitter BS$_2$
(\mbox{$R_2$ $\! \neq$ $\! 0$}), the resonance of
the double-cavity system is split (quasi-Rabi splitting), so that now 
\mbox{$|G_{11}|$ $\! \approx$ $\! 0$} for \mbox{$k$ $\!=$
$\!k_0$ $\!\pm$ $\!\Delta k$}, where 
\begin{eqnarray}
 \Delta k \approx \sqrt{\frac{R_2}{L_H L_V}} \,.
 \label{deltak}\end{eqnarray}
In the middle of the interval between the two resonant wave-number
values, the system is transparent, i.e., \mbox{$G_{11}$
$\! \approx$ $\! 1$} for \mbox{$k$ $\!=$ $\!k_0$}.

Let us consider, for a moment, the isolated system consisting
of the two cavities that are decoupled from the external world
(\mbox{$R_1$ $\!=$ $\!0$}), but coupled to each other
(\mbox{$R_2$ $\!\neq$ $\!0$}). Solving (for nonabsorbing devices)   
the eigenmode equation, we find that the symmetric and
antisymmetric combinations of the modes of the individual
cavities form modes of the double-cavity system which
just differ in the value of $\Delta k$ given by
Eq.~(\ref{deltak}). Hence, an incoming wave of wave number $k$ 
can be regarded as being coupled by the beam splitter BS$_{1}$
to these two modes. 
Whereas for \mbox{$k$ $\!=$ $\!k_0$ $\!\pm$ $\!\Delta k$}
one of the two modes can be excited, the couplings to them
interfere destructively for \mbox{$k$ $\!=$ $\!k_0$}, so that
the incoming field is virtually decoupled from the cavities and
the system becomes transparent. 

The behavior of a double-cavity resonator resembles 
the behavior of several other physical systems (for details,
see Appendices \ref{App-EIT} -- \ref{App-Zeno}). The system
has many similarities to a three-level EIT medium
(Appendix \ref{App-EIT}) and to a one-dimensional atom in a
cavity (Appendix \ref{App-1D-atom}), and there is a close
relationship to the so-called interaction-free-measurement
and the quantum-Zeno effect (Appen\-\mbox{dix \ref{App-Zeno})}.


\subsection{Time delay}
\label{Sec-delay}

{F}rom the input-output relation (\ref{inout}) it follows that
outgoing and incoming pulses in the time domain are related
to each other as
\begin{equation}
\label{inout1}
a_{out}(t) 
= \int {G}(\tau)\,a_{in}(t\!-\!\tau)\, d\tau,
\end{equation}
where
\begin{equation}
\label{G11time}
{G}(\tau) = \frac{1}{2\pi}\int
e^{-i\omega \tau} G_{11}(\omega)\,d\omega,
\end{equation}
with $G_{11}(\omega)$ from Eq.~(\ref{G11}).
When absorption can be disregarded \mbox{($A_{l},A_{Mm}$
$\!=$ $\!0$)} and the reflection coefficients $R_{l}$
are small, then $G_{11}(\omega)$ can be expanded around
\mbox{$\omega$ $\!=$ $\!\omega_0$} to obtain
\begin{eqnarray}
G_{11}(\omega) &\approx& 
\left(
1 + i L_D \,\delta k -  L_D^2 \,\delta k^2 \right)
\nonumber\\[.5ex]
 &\approx& 
 \exp\!\left(i L_D \,\delta k -
 {\textstyle\frac{1}{2}}L_D^2 \,\delta k^2\right)
\label{G11delay2}
\end{eqnarray}
\mbox{[$\delta k$ $\!=$ $\!(\omega$ $\!-$ $\!\omega_0)/c$
$\!\ll$ $1/L_D$]},
where
\begin{eqnarray}
L_D = \frac{R_1 L_H}{2R_2} \,.
\end{eqnarray}

Let us consider an incoming pulse that has a
mid-fre\-quency \mbox{$\omega_0$ $\!=$ $\!k_0 c$} and a spectral
width smaller than $c/L_D$. In that case, the response function
$G(\tau)$ in Eq.~(\ref{G11time}) can be obtained from 
Eq.~(\ref{G11delay2}) with the approximate expression of
$G_{11}(\omega)$ given in Eq.~(\ref{G11delay2}):
\begin{eqnarray}
G(\tau) \approx \frac{1}{\sqrt{2\pi}\tau_D}
\,\exp\!\left[-\frac{(\tau-\tau_D)^2}{2\tau_D^2}\right]  ,
\label{G11time1}
\end{eqnarray}
where   
\begin{eqnarray}
 \tau _D = \frac{L_D}{c} = \frac{R_1 L_H}{2R_2 c} .
 \label{tauDapprox}
\end{eqnarray}
Comparing with Eq.~(\ref{inout1}), we see that
the outgoing pulse is delayed by  $\tau_D$ and its envelope  
is broadened by the same amount $\tau_D$ (i.e., the envelope of the power 
intensity is broadened by $\tau_D/\sqrt{2}$).


\subsection{Internal fields}
\label{Sec-intens}

\begin{figure}[!t]
\noindent
\centerline{\epsfig{figure=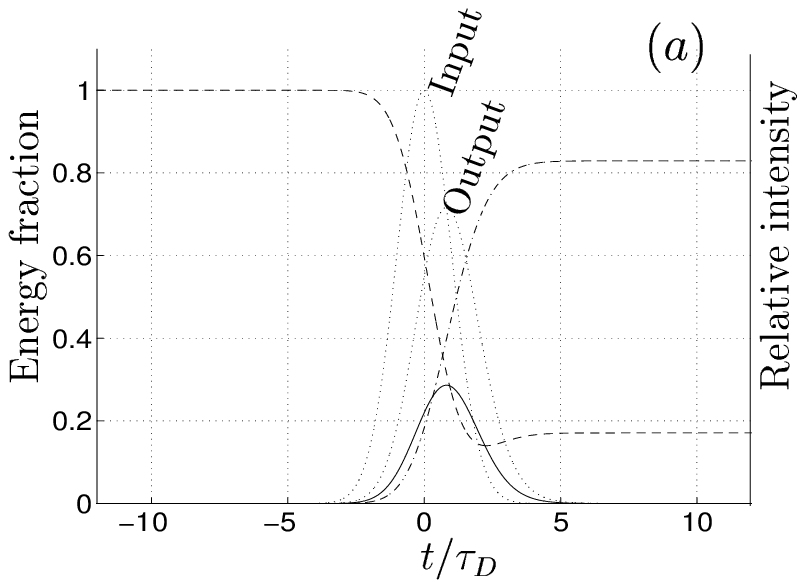,width=0.9\linewidth}}
\\[1ex]
\centerline{\epsfig{figure=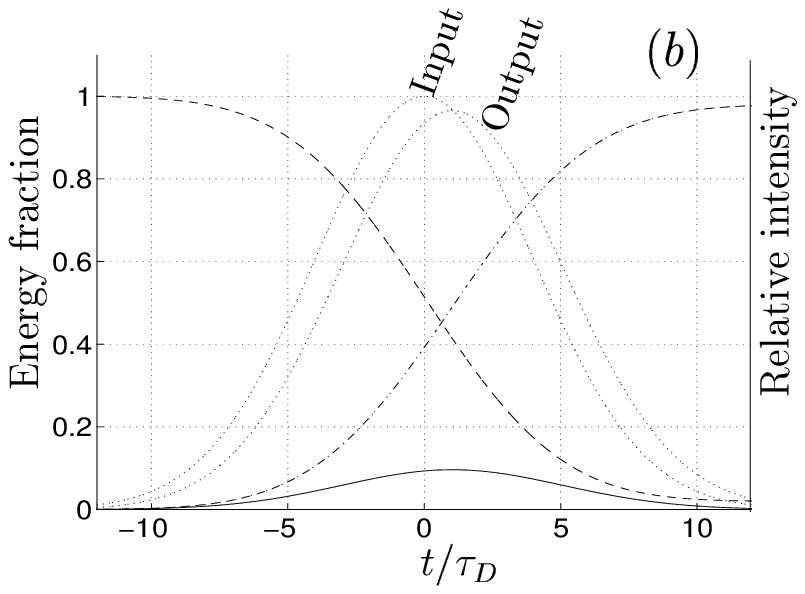,width=0.9\linewidth}}
\caption{ 
Time evolution 
of the energy fraction in the space in front of the system (broken
line), inside the horizontal cavity (full line), and in the space behind the
system (dash-dotted line),
and relative intensities entering and leaving the system at BS$_1$ (dotted
lines). The input pulse is a Gaussian with a half width $\tau_D$ (a), and
$4\tau_{d}$ (b), respectively.
 The cavity parameters  are 
$L_V = 120\lambda$, $L_H=30\lambda$,  $R_1=0.1$, 
$R_2=10^{-6}$, no absorption.
}
\label{figtemp}
\end{figure}

The complex amplitudes (in the frequency domain) of the fields
inside the double-cavity system are given in
Appendix~\ref{Ap-intracavity}. Expanding the amplitudes
of the fields in the horizontal cavity, Eqs.~(\ref{Ap2M1}),
(\ref{ApM12}), (\ref{Ap2M3}), and (\ref{ApM32}), around
\mbox{$\omega$ $\!=$ $\!\omega_0$} we find, on neglecting
again losses, that $-a_{2,M1}$ $\!\approx$ $\!a_{M1,2}$ $\!\approx$
$\!-a_{2,M3}$ $\!\approx$ $\!a_{M3,2}$ $\!\equiv$ $a_H(\omega)$, where
\begin{equation}
 a_{H}(\omega) = \frac{1}{2}\sqrt{\frac{R_1}{R_2}}
 \left( 1+i \delta k L_D - \delta k ^2 L_D^2 \right) a(\omega) 
 \label{cH}
\end{equation}
\mbox{($b$ $\!=$ $\!0$)}.
We compare Eq.~(\ref{cH}) with Eq.~(\ref{G11delay2}) and 
see that
\begin{equation}
\label{cH1}
 a_{H}(\omega) \approx \frac{1}{2}\sqrt{\frac{R_1}{R_2}}\,
 G_{11}(\omega)\, a(\omega)
\end{equation}
\mbox{($\delta k$ $\!\ll$ $1/L_D$)}.
Thus, the resonant field in the horizontal cavity has an intensity
that is $R_1/(4R_2)$ times larger than that of the incoming field.
In the time domain, it is obviously a pulse that is   
delayed and broadened by $\tau_D$, 
i.e., 
\begin{equation}
\label{cH1time}
 a_{H}(t) \approx \frac{1}{2}\sqrt{\frac{R_1}{R_2}}\,
  \int {G}(\tau)\,a_{in}(t\!-\!\tau)\,d\tau.
\end{equation}

In the same approximation, from Eqs.~(\ref{Ap12}),
(\ref{Ap1M4}), (\ref{ApM41}), (\ref{Ap2M2}), and (\ref{ApM22})
we find that the leading term of the field amplitudes in the vertical
cavity is \mbox{$a_{V}(\omega)$ $\!=$ $ia(\omega)\sqrt{R_1}/2$}.
That is, the intensity of the field in the vertical cavity is
only a fraction of $R_1/4$ of the intensity of the incoming field.
Since noticeable XPM in the probe will require a large field in the
horizontal cavity, it is crucial to have $R_2$ as
small as possible.

The temporal evolution of field energies on the left-hand side
of the system (incoming and reflected field), in the horizontal
cavity, and on the right-hand side of the system (transmitted field),
as well as the input and output field intensities
are illustrated in Fig.~\ref{figtemp}. 
One can see the importance of a proper choice of the signal pulse length.
For short pulses (relative to
$\tau_D$), strong electric intensity develops in the horizontal cavity, but the
output pulse is deformed with a large fraction 
of the pulse being reflected. Long pulses keep
their shape, but cannot produce very strong intensities.


\subsection{Influence of losses}
\label{Sec-losses}

Let us comment on the
conditions under which the material absorption
of the mirrors and the beam splitters may be disregarded. 
{F}rom an expansion of the expressions in
Eqs.~(\ref{G11}) -- (\ref{G22}) it follows that when
\begin{eqnarray}
 A_{1}, A_{M2}, A_{M4}  &\ll& R_1\,, 
 \label{loss1}
 \\
 A_{2}, A_{M1}, A_{M3}  &\ll& \frac{R_2}{R_1}\,,  
 \label{loss2}
\end{eqnarray}
then the absorption coefficients can be neglected.
Since $R_2$ should be as small as possible, the condition
(\ref{loss2}) is the most restrictive. 

The probabilities of absorption of monochromatic photons 
are given by \mbox{$P_a$ $\!=$ $\!1-$ $\!|G_{11}|^2$ $\!-$
$\!|G_{12}|^2$}
and 
\mbox{$P_b$ $\!=$ $\!1$ $\!-$ $\!|G_{22}|^2$ $\!-$ $\!|G_{21}|^2$}.
In particular, for small absorption we have 
\mbox{$P_a$ $\!\approx$ $\!P_b$ $\!\equiv$ $\!P$},
and if $k$ is close to $k_0$, expansion yields
\begin{eqnarray}
\lefteqn{
 P 
 \approx  A_{1} + A_{M4}\frac{R_1}{4}
} 
\nonumber \\&&\hspace{2ex}
 +\left( A_{M1}+A_{M3}+A_{2} \right)
 \frac{R_1}{4R_2} \left( 1 - \frac{R_1^2}{4} \right)
\nonumber \\&&\hspace{2ex}
 + \,\delta k^2 L_D^2 
 \left[ \frac{2A_1+A_{M2}+A_{M4}}{R_1} \right.
\nonumber \\&&\hspace{2ex}
 \left. -\left( A_{M1}+A_{M3}+A_2 \right) \frac{R_1}{4R_2}
 \left(1 + R_1\right)
 \right].
 \label{absprobab}
\end{eqnarray}
To obtain the absorption probability of a photon associated with
a wavepacket, one has to average over the corresponding
$k$\,distribution.

\begin{figure}
\centerline{\epsfig{figure=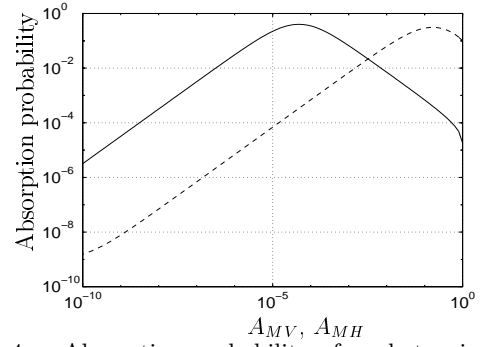,width=0.8\linewidth}}
\caption{ 
Absorption probability of a photon in a Gaussian wave packet with a 
temporal half-width of $\tau_{D}$ in dependence on the mirror absorption
coefficients. Full line: $A_{M2}=A_{M4}=0$, $A_{M1}=A_{M3}=A_{MH}$;
dashed line: $A_{M1}=A_{M3}=0$, $A_{M2}=A_{M4}=A_{MV}$. The other cavity
parameters are $L_H=30\lambda$,  $L_V=120\lambda$, $R_1=10^{-1}$, 
$R_{2}=10^{-6}$, $A_1=A_2=0$.
}
\label{figabs}
\end{figure}

Figure \ref{figabs} illustrates
the dependence of the absorption probability on the 
mirror absorption coefficients for a Gaussian wave packet
of the temporal half-width $\tau_D$.
[Note that for such a wave packet the second-order expansion
in $\delta k$ as given in Eq.~(\ref{absprobab}) fails,
because the $k$\,distribution of the wave packet is too
broad.] It is seen that the absorption in the horizontal cavity (full
line) plays the dominant role. The decrease of the absorption
probability for large absorption coefficients
(right to the knee of the full line)
corresponds to the interaction-free measurement effect
(Appendix \ref{App-Zeno}).


\subsection{Low reflectivity beam-splitters}

As already pointed out, the reflectivity of the second
beam splitter, $R_2$, should be as small
as possible. To achieve this goal, we first assume that it
is made from a thin dielectric plate of thickness $d$
\mbox{($d$ $\!\ll$ $\!\lambda$)} and refractive index 
$n$ (the refractive index of the medium be $\approx 1$), so that
\begin{eqnarray}
R_2 = r_2^2 \approx \left[\frac{n^2-1}{\sqrt{2}}\,kd\right]^2.   
 \label{reflectivity}
\end{eqnarray}
For \mbox{$d$ $\!\approx$ $\!10$\,nm} 
\mbox{($kd$ $\!\approx$ $\!10^{-2}$)}, $R_2$ can thus be estimated
to be \mbox{$R_2$ $\!\approx$ $\!10^{-4}$}, which will be not
small enough. Let us anticipate the progress of nanotechnology,
assuming the possibility to built very thin and sparse
dielectric nanostructures, so that \mbox{$R_2$ $\!\approx$
$\!10^{-6}$} becomes possible. This value would be realized by
a dielectric plate of $1$\,nm thickness.  Another possibility 
would be constructing a sparse network-like structure of
$10$\,nm fibers, which would cover about 10\%
of the (otherwise empty) beam-splitter area.

One could also think about a cavity
made of some solid material. Then,
including in it a weakly reflecting beam splitter  
would not be difficult, because it could be realized by
a thin layer of material with slightly different index of
refraction. A disadvantage may be
the somewhat increased 
absorption losses.
Although EIT has successfully been observed in doped
solid materials \cite{EITsolid}, in such materials a 
smaller achievable XPM than in gases is expected, because 
of the lower dipole moments of the doping atoms.


\section{Conditional phase shift}
\label{se3}


A photonic qubit can be realized by a single photon which can travel
along two alternate paths (e.g., two  branches of an interferometer),
the corresponding states being denoted as $|0\rangle$ and $|1\rangle$. 
Let us consider two qubits
such that in the state $|11\rangle$ the paths of the two
photons partly overlap in a Kerr nonlinear medium. Further,
let the nonlinearity be so strong that the presence of
one photon changes the refractive index
appreciated by the other photon
in order to introduce a $\pi$ phase shift. 
The state  $|11\rangle$ thus transforms as
\mbox{$|11\rangle$ $\!\to$ $\!-$ $\!|11\rangle$}, while
the other states remain unchanged: 
\mbox{$|00\rangle$ $\!\to$ $\!|00\rangle$}, 
\mbox{$|01\rangle$ $\!\to$ $\!|01\rangle$}, and
\mbox{$|10\rangle$ $\!\to$ $\!|10\rangle$}.
Such a transformation, accompanied with single qubit
transformations (which are trivial for the photonic
qubit realization considered), can serve as a building block
for quantum computation (see, e.g., \cite{Fortphys}), 
or for achieving quantum teleportation \cite{ParisVitali}.


\subsection{EIT medium}
\label{se3.1}

\begin{figure}[!t]
\noindent
\centerline{\epsfig{figure=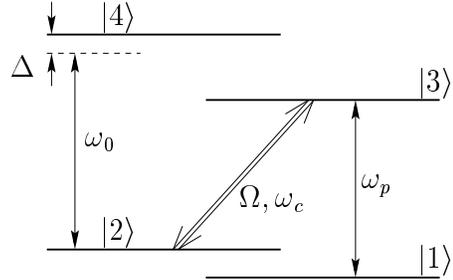,width=0.7\linewidth}}
\caption{ 
Four-level scheme of the EIT medium.
}
\label{atschem}
\end{figure}

To realize the desired large XPM, let us combine  the action of the
double-cavity system and that of a four-level EIT medium of the type studied in
Ref.~\cite{Schm96} (Fig.~\ref{atschem}). We assume an atomic medium with the
four-level structure as in  Fig.~\ref{atschem} whose transition $|2\rangle
\leftrightarrow |3\rangle$ is driven by an external coupling field of Rabi
frequency $\Omega$.  The probe field (mid-frequency $\omega_p$)
is resonant with the  $|1\rangle \leftrightarrow |3\rangle$
transition, whereas the signal field (mid-frequency \mbox{$\omega_s$
$\!=$ $\!\omega_0$})
is detuned by $\Delta$ from the $|2\rangle
\leftrightarrow |4\rangle$ transition.
Using fourth-order perturbation theory, one can find that the
Kerr index of refraction felt by the probe is 
\begin{eqnarray}
  n_{K} = \frac{N \mu_{13}^2 \mu_{24}^2}
 {8\epsilon_{0} \hbar^{3} |\Omega|^2 \Delta}
 |E_s|^2 , 
 \label{n2eq}
\end{eqnarray}
where $N$ is the atomic density, $\mu_{13}$ and  $\mu_{24}$ are
respectively the dipole matrix elements of the transitions
\mbox{$|1\rangle$ $\!\leftrightarrow $ $\!|3\rangle$} and
\mbox{$|2\rangle$ $\!\leftrightarrow $ $\!|4\rangle$},
and $E_s$ is the electric field strength of the signal.
The group velocity $v_g$ of the probe can be given by \cite{Hau99}
\begin{eqnarray}
v_{g} = c
\left[n_0 + \omega_p \frac{dn_0}{d\omega_p}\right]^{-1}
\approx \frac{\epsilon_0 \hbar \lambda |\Omega|^2}{4\pi N \mu_{13}^2}\,,
\label{vg}
\end{eqnarray}
where $n_0$ is the signal-independent part of the refractive
index.

During the propagation in the EIT medium,
the signal and the probe suffer from several kinds of
losses. The most important losses result from the finite
frequency window for the probe
(frequency components outside a very narrow interval
with respect to the  
\mbox{$|1\rangle$ $\!-$ $\!|3\rangle$} transition are
partially absorbed),
and two photon losses due
to the simultaneous absorption of probe and signal photons 
(exciting the atomic state $|4\rangle$).
The loss coefficient $\alpha_1$ (inverse absorption length) for 
single-photon probe absorption
can be given by \cite{Schm98,Harris99}
\begin{eqnarray}
\alpha_1 = \frac{32 \pi^2 N \mu_{13}^2 \gamma_{3} \delta^2}
 {\epsilon_0 \hbar \lambda |\Omega|^4} \,,
 \label{alpha1}
\end{eqnarray}
where $\delta$ is the detuning of the probe from the
\mbox{$|1\rangle$ $\!-$ $\!|3\rangle$} transition, and
$\gamma_{3}$ is the decay rate of the 
state $|3\rangle$.
The loss coefficient $\alpha_2$ for the (simultaneous)
two-photon absorption  can be written in
the form of \cite{Schm98})
\begin{eqnarray}
 \alpha_2 = \frac{\pi^2 N \mu_{13}^2\mu_{24}^2 \gamma_{4}}
 {2 \epsilon_0 \hbar^3 \lambda |\Omega|^2 \Delta^2} |E_s|^2\, ,
 \label{alpha2}
\end{eqnarray}
where $\gamma_{4}$ is the atomic decay rate of the 
state $|4\rangle$.

\subsection{Conditional single-photon phase shift}
\label{se3.2}

Let us assume that the signal and the probe are sent,
according to Fig.~\ref{fig1}, to the double-cavity system
complemented with the EIT medium. The conditional phase shift
$\Delta \phi$ of the probe due to XPM is
\begin{eqnarray}
 \Delta \phi = \int   n_K  k_p dl \, ,
\end{eqnarray}
where
the integral runs over the propagation length of the probe.
For a pulse propagating with the group velocity 
$v_g$ in an otherwise homogeneous medium we may write
$dl = v_g dt$ and use Eqs. (\ref{n2eq}), (\ref{vg}) so that
\begin{eqnarray}
 \Delta \phi = \frac{\mu_{24}^2}{16 \hbar^2 \Delta}
 \int |E_s(t)|^2 dt \, .
 \label{Dphit}
\end{eqnarray}
The integration runs over the time during which the probe pulse is inside
the medium.
Let the input signal be a single-photon
Gaussian pulse of half-width $\tau_s$, i.e., its electric field reads 
\begin{eqnarray}
  E_s^{\rm (in)}=E_0 \exp\!\left(-\frac{t^2}{4\tau_s^2}-i\omega_0 t\right) , 
  \label{Etime}
\end{eqnarray}
where
\begin{eqnarray}
 E_0^2 = \sqrt{\frac{2}{\pi}} \frac{\hbar \omega_0}
 {c \epsilon_0 S \tau_s}\,,
\end{eqnarray}
with $S$ being the cross-sectional area of the pulse.
We apply the transformation (\ref{cH1time}) to get the amplitude of the
intracavity field propagating from BS$_2$ to the mirrors of the horizontal
cavity in the form
\begin{eqnarray}
 E_{s,H}= \frac{E_0}{2} \sqrt{\frac{R_1}{R_2}} 
 \left(1+ \frac{\tau_D^2}{2\tau_s^2}\right)^{-1/2}
 \nonumber \\
 \times
 \exp\!\left[ -\frac{(t-\tau_D)^2}
 {4\tau_s^2 \left( 1+\frac{\tau_D^2}{2\tau_s ^2} \right)}
 -i \omega_0 t \right].
\end{eqnarray}
This field combines with the reflected fields to a standing wave whose squared
amplitude averaged along $L_H$ is \mbox{$|E_s(t)|^2$ $\!=$
$\!2|E_{s,H}|^2$}. 
When the probe group velocity is sufficiently low so
that the time which the probe spends in the medium is longer
than the time duration of the signal in the cavity,
\begin{eqnarray}
 L/v_g \gtrsim 4 \,\sqrt{\tau_s^2 +\tau_D^2/2}\,,
 \label{Lvg}
\end{eqnarray}
then the time integral in Eq.~(\ref{Dphit}) can be approximated by
integration in infinite limits as
\begin{eqnarray}
 \int |E_s(t)|^2 dt = \frac{R_1}{R_2} \frac{2\pi \hbar}
 {\epsilon_0 \lambda S} \left(1+\frac{\tau_D^2}{2\tau_s^2}\right)^{-1/2} .
 \label{E2int}
\end{eqnarray}
Thus, from (\ref{E2int}) and (\ref{Dphit}) the 
the phase shift of the probe is estimated to be  
\begin{eqnarray}
 \Delta \phi \approx \frac{\pi}{8}
 \frac{R_1}{R_2} 
 \frac{\mu_{24}^2}{\epsilon_0 \hbar \lambda S \Delta}
 \left(1+\frac{\tau_D^2}{2\tau_s^2}\right)^{-1/2}.
 \label{Dphires}
\end{eqnarray}


\subsection{Conditions for the parameters}
\label{se3.3}

{F}rom Eq.~(\ref{Dphires}) it is seen that the duration of
the signal pulse, $\tau_s$, should be as long as possible,
but at least comparable with $\tau_D$. On the other hand,
$\tau_s$ should be limited by the transmission time
of the probe according to the condition (\ref{Lvg}).
Otherwise, the full signal would not be used.
Equation (\ref{Dphires}) further reveals that the phase shift 
does not explicitly depend on the parameters $N$, $\mu_{13}$,
and $|\Omega|$. However, these parameters must be chosen such
that the group velocity (\ref{vg}) is low enough to
satisfy the condition (\ref{Lvg}).  Assuming
\mbox{$L$ $\!\approx$ $\!L_H$} and
\begin{eqnarray}
 \left(1+\frac{\tau_D^2}{2\tau_s^2}\right)^{-1/2} \approx \frac{1}{2}\,,
\end{eqnarray}
we get from Eqs.~(\ref{Lvg}) and (\ref{tauDapprox}) the
following condition for the group velocity:
\begin{eqnarray}
 v_g \lesssim \frac{\sqrt{6}}{8}\,\frac{R_2}{R_1}\, c.
 \label{vgRc}
\end{eqnarray}
Recalling Eq.~(\ref{vg}), we see that the Rabi frequency of
the driven transition
must satisfy
the condition
\begin{eqnarray}
|\Omega|^2  \lesssim \frac{\sqrt{6}\pi}{2}
\,\frac{R_2}{R_1}\,\frac{c \mu_{13}^2}{\epsilon_0 \hbar \lambda}\, N.
 \label{omegacond}
\end{eqnarray}

Further, the parameters should also be chosen such that
the losses are sufficiently small.
The probability of two-photon absorption,
\begin{eqnarray}
 P_2 \approx \int dl\,\alpha_2 (|E_s|^2) = 
 v_g \int dt\,\alpha_2 (|E_s|^2),
\end{eqnarray}
can be calculated with the help of Eqs.~(\ref{vg}), (\ref{alpha2}),
and (\ref{E2int}). The condition that \mbox{$P_2$ $\!\ll$ $\!1$}
then reads as 
\begin{eqnarray}
 P_2 \approx
 \frac{\pi^2}{4} \frac{\mu_{24}^2 \gamma_4}
 {\epsilon_0 \hbar S \lambda \Delta ^2} \frac{R_1}{R_2}
 \left(1+\frac{\tau_D^2}{2\tau_s^2}\right)^{-1/2}
 \ll 1,
 \label{condP2}
\end{eqnarray}
and the condition of the two-photon
absorption
probability being much smaller than the
phase shift (\ref{Dphires}) is
\begin{eqnarray}
 \frac{P_2}{\Delta \phi} \approx \frac{2\pi \gamma_4}{\Delta} \ll 1 ,
 \label{P2Dphi}
\end{eqnarray}
which corresponds to the result in Ref.~\cite{Schm96}.
The condition of negligible single-photon absorption of the probe
\mbox{$\alpha_1 L$ $\!\ll$ $\!1$} can be written as,
on recalling Eq.~(\ref{alpha1}),
\begin{eqnarray}
\frac{32 \pi^2 N L \mu_{13}^2 \gamma_{3} \delta^2}
 {\epsilon_0 \hbar \lambda |\Omega|^4} \ll 1. 
 \label{conddelta1}
\end{eqnarray}

We have assumed that
during the interaction with the signal,
the whole probe pulse is inside the medium.
Thus, the time duration
\mbox{$\tau_p$ $\!\approx$ $\!\delta^{-1}$} of the probe
must be shorter than 
its propagation time through the medium.
{F}rom Eq.~(\ref{vg}) it then follows that this assumption
yields the condition that
\begin{eqnarray}
\frac{1}{\delta} \ll \frac{L}{v_g} = \frac{4\pi NL\mu_{13}^2}
 {\epsilon_0 \hbar \lambda |\Omega|^2} \,.
 \label{conddelta2}
\end{eqnarray}
By combining the conditions (\ref{conddelta1}) and (\ref{conddelta2}),
we obtain a condition for the atomic density, 
\begin{eqnarray}
 N \gg \frac{2 \epsilon_0 \hbar \lambda \gamma_3}{\mu_{13}^2L}\,,
 \label{Ncondition} 
\end{eqnarray}
and a condition for the line-width of the probe,
\begin{eqnarray}
 \delta \ll \frac{|\Omega|^2}{8\pi \gamma_3}\,.
\end{eqnarray}

{F}rom Eq. (\ref{Dphires}) it follows that the ratio $R_1/R_2$ should be chosen
as large as possible. This would suggest using large $R_1$, but 
such a choice would mean large probability of reflection of the signal photon. 
In particular, expansion of $G_{11}$ and $G_{21}$ for resonant signal yields
$G_{11} \approx 1-R_1^2/8$  and $G_{21} \approx R_1/2$. We can see that using
$R_1 \approx 0.1$ is a reasonable choice.


\subsection{Experimental parameters}
\label{se3.4}
 
In the experimental demonstration of ultraslow group
velocity in Ref.~\cite{Kash99} a gas of rubidium atoms
was used.
In this case, \mbox{$\lambda$ $\!=$ $\!795$\,nm},
\mbox{$\mu_{13}$ $\!\approx$ $\!\mu_{24}$ $\!\approx$
$\!10^{-29}$\,Cm}, and
\mbox{$\gamma_3$ $\!\approx$ $\!\gamma_4$ 
$\!\approx$ $\!10^6$\,s$^{-1}$}.
Let the signal beam diameter be \mbox{$\approx$
$\!L_H$ $\!\approx$ $\!10$\,$\mu$m} so that
\mbox{$S$ $\!\approx$ $\!10^{-10}$\,m$^{2}$}.
The condition (\ref{P2Dphi}) is satisfied,
\mbox{$P_2/ \Delta \phi$ $\!\approx$ $\!0.1$}, if
\mbox{$\Delta$ $\!=$ $\!10^{8}$\,s$^{-1}$}.
{F}rom Eq.~(\ref{Dphires}) it then follows that for
\begin{eqnarray}
  \frac{R_1}{R_2} \approx 10^{5}
  \label{R12} 
\end{eqnarray}
a phase shift of \mbox{$\Delta\phi$ $\!\approx$ $\!\pi$}
can be achieved. 

The condition (\ref{Ncondition}) requires
that \mbox{$N$ $\!\gg$ $\!10^{12}$\,cm$^{-3}$}. Let
us assume an atomic density of
\mbox{$N$ $\!\approx$ $\!10^{14}$\,cm$^{-3}$}.
Since the detuning $\Delta$ is very small, the gas has to
be laser-cooled to avoid the Doppler broadening.
The condition (\ref{omegacond}) then implies that
the Rabi frequency of the driven atomic transition should
be \mbox{$|\Omega|$ $\!\approx$ $\!10^{9}$\,s$^{-1}$}.
With these values we find from Eqs.~(\ref{n2eq})
and (\ref{vg})
for the group velocity \mbox{$v_g$ $\!\approx$ $\!10^3$\,ms$^{-1}$},
the switching time being 
\mbox{$L_H/v_g$ $\!\approx$ $\!10$\,ns}.
According to the condition (\ref{conddelta2}), the
length of the probe pulse should chosen such that
\mbox{$\tau_p$ $\!\gg$ $\!10^{-11}$\,s}. Choosing \mbox{$\tau_p$
$\!\approx$ $\!1$\,ns} (i.e., \mbox{$\delta$ $\!\approx$
$\!10^{9}$\,s$^{-1}$}),
Eq.~(\ref{alpha1}) yields \mbox{$\alpha_1$ 
$\!\approx$ $\!10^{4}$\,m$^{-1}$}, so that the linear absorption
is relatively small \mbox{($\alpha_1L$ $\!\approx$ $\!0.1$, 
$L \approx L_H$)}. 

The condition (\ref{R12}) is a strong requirement for building the
subtle beam-splitter needed; for \mbox{$R_1$ $\!\approx$ $\!10^{-1}$} one
needs \mbox{$R_2$ $\!\approx$ $\!10^{-6}$}. With respect
to the condition (\ref{loss2}), we see that the absorption
coefficients of the mirrors should be at least one order of
magnitude smaller than the ratio $R_2/R_1$. 
Thus the quality of presently available mirrors
(absorption $\approx$  10$^{-6}$ \cite{Kuhn}) is 
nearly
sufficient.


\subsection{Comparison with free-medium scheme}
\label{se3.5}

Let us compare the performance of the coupled-cavity
scheme with the free-medium scheme in Ref.~\cite{Lukin00}.
Assuming a diameter of \mbox{$l_D$ $\!\approx$ $\!10\,\mu$m}
of the  copropagating signal and probe and a pulse linewidth of
\mbox{$\Delta \omega$ $\!\approx$ $\!1$\,MHz}, the signal field
would produce
a shift of the refraction index for the probe of the order
\mbox{$n_K$ $\!\approx$ $\!10^{-4}$}. Thus, in the
free-medium scheme a propagation length of
\mbox{$l_0$ $\!\approx$ $\!1$\,cm} would be necessary to achieve a
phase shift of the order of $\pi$, which implies
a switching time of \mbox{$l_0/v_g$ $\!\approx$ $\!1$\,ms}. This
is about $10^5$ times longer than the switching time
in the coupled-cavity scheme.
Note that the length $l_0$ is about $10^2$ times longer than
the Rayleigh distance
\mbox{$l_R$ $\!\approx$ $\!l_D^2/\lambda$ $\!\approx$ $\!100\,\mu$}
(see, e.g., \cite{Teich}). Thus, a sophisticated
waveguide structure or a system of refocusings would be necessary
which makes the free-medium approach technically demanding as well,
so that other possibilities of confining the interacting beams 
have been of interest \cite{Lukin-private}.


\section{Conclusion}
\label{se4}

A suitable combination of optical resonators with an EIT medium
can increase the XPM effect to be used in quantum information
processing. The cavity system 
emulates the EIT response for the signal: the 
{\em signal\/} is slowed down and
simultaneously its intensity is enhanced. The {\em probe\/} 
is not influenced by the
cavity, but it propagates in the EIT medium whose properties
are determined by the signal. 
By coupling the cavities with weakly reflecting beam splitters rather than 
with
weakly transmitting mirrors, large losses can be avoided. 
This makes the system especially attractive for applications in quantum
information processing.

To achieve a large ($\approx \pi$) phase-shift by means of a
single signal photon, several technical problems must be solved.
As the most important, very low-reflectivity \mbox{($R$
$\!\lesssim$ $\!10^{-6}$)} beam-splitters must be available
and a suitable method for keeping the 
cooled 
EIT atoms in an optical cavity
must be developed.
The device then could serve as a relatively fast
(\mbox{$\approx$ $\!10$\,ns} switching time) quantum logic gate.


\acknowledgments

This work was supported by the Deutsche Forschungsgemeinschaft. 
T.O. is grateful to H. Bartelt, M. Fleischhauer, M.D. Lukin,
A. Kuhn, and S. Scheel for discussions.

\appendix


\section{Linear response of a conventional cavity}
\label{Ap-cavity1}

Consider a cavity constructed from two parallel mirrors separated
by length $L$. Assume the transformation rule for the mirrors as
\begin{eqnarray}
 \left(  
 \begin{array}{c}
  a' \\  b'
 \end{array}
 \right) =  \left(
 \begin{array}{rr}
  it & -r \\
  -r & it
 \end{array}
  \right) 
  \left( 
 \begin{array}{c}
   a \\  b
 \end{array}  
 \right) ,
\end{eqnarray}
where $a$,\,$b$ are the input field amplitudes, $a'$,\,$b'$ are
the output field amplitudes, \mbox{$r^2$ $\!\equiv$ $\!R$
$\!\approx$ $\!1$} is the mirror reflectivity, 
\mbox{$t^2$ $\!\equiv$ $\!T$ $\!\ll$ $\!1$} is the
transmissivity, and \mbox{$R$ $\!+$ $\!T$ $\!+$ $\!A$ $\!=$
$\!1$}, with $A$ being the absorption coefficient.
When a unit input field of wave number $k$ is fed into
the cavity, then the amplitude of the output field that is
generated by reflection at the entrance mirror is
\begin{eqnarray}
 q= r \frac{(1-A) e^{i2kL} -1}{1-Re^{i2kL}}\,,
\end{eqnarray}
and the amplitude of the (transmitted) output field at
the other mirror is 
\begin{eqnarray}
 p= -\frac{T e^{ikL}}{1-Re^{i2kL}}\,.
\end{eqnarray}
Close to resonance, \mbox{$k$ $\!\approx$ $\!k_0$}
(\mbox{$k_0 L$ $\!=$ $\!n\pi$}), the
transmission through the cavity
is then
\begin{eqnarray}
 {\cal T} \equiv |p|^2 \approx \frac{T^2}{(T+A)^2} \frac{1}
 {1+(k-k_0)^2/\Delta k^2}\,,
\end{eqnarray}
and the reflection at the cavity
reads
\begin{eqnarray}
\lefteqn{
 {\cal R} \equiv |q|^2 \approx \frac{A^2 R}{(T+A)^2} \frac{1}
 {1+(k\!-\!k_0)^2/\Delta k^2}
}
\nonumber \\ &&
 +\,(1-A) \frac{(k\!-\!k_0)^2}{\Delta k^2} \frac{1}
 {1+(k\!-\!k_0)^2/\Delta k^2} \,,
\end{eqnarray}
where the line-width $\Delta k$ is given by
\begin{eqnarray}
 \Delta k = \frac{T+A}{2\sqrt{R}L}.
\end{eqnarray}
Thus, if \mbox{$A$ $\!\approx$ $\!T$}, we get 
\mbox{${\cal T}$ $\!\approx$ $\!1/4$} and
\mbox{${\cal R}$ $\!\approx$ $\!R/4$} in the
resonance regime.


\section{Analogy between coupled resonators and EIT media}
\label{App-EIT}

The action on light of the double-cavity system is
analogous to that of a three-level EIT medium (see the
level scheme in Fig.~\ref{atschem}, without 
$|4\rangle$). The empty double-cavity system corresponds
to the atomic (ground) state $|1\rangle$ and light in the
vertical cavity corresponds to the excited atomic state
$|3\rangle$. Accordingly, light in the horizontal cavity
corresponds to the (auxiliary) atomic state $|2\rangle$.   
Coupling of the atomic states $|3\rangle$ and $|2\rangle$,
which gives rise to the dressed (Rabi-split) states,
is analogous to the cavity coupling by the beam-splitter BS$_2$.

{F}rom Fig.~\ref{fig2} it is seen that the
linear response of the cavity system to the incoming light
resembles
the response of an EIT medium.
Physically, the two systems are of course quite different.
In particular, 
frequency components that are slightly off-resonant are absorbed
by the EIT medium but are reflected by the cavity system.

Recently large phase shifts
in moving media with EIT-type dispersion
have been predicted \cite{Ulf}. 
A corresponding effect is observed in the double-cavity system
when it is moved along the signal-propagation direction. 
The produced phase shift is \mbox{$\Delta \phi_{v}$ $\!=$
$\!\omega_0\tau_D v/c$}, where $v$ is the velocity of the
system. It can be large if $\tau_D$ [Eq.~(\ref{tauDapprox})]
is sufficiently large, even when the motion of the system
is relatively slow.


\section{Analogy between coupled resonators and one-dimensional atoms}
\label{App-1D-atom}

The double-cavity system is also analogous to 
a {\em one-dimensional atom\/} in the bad-cavity regime
\cite{Turch95}. Here a two-level atom inside a cavity is resonantly
coupled to the cavity field such that the Rabi splitting is larger than
the spontaneous decay rate of the atom, whereas the decay rate
of the cavity field is larger than
the Rabi splitting.
In the double-cavity system, the horizontal cavity plays the role of
the atom, and the vertical cavity plays the role of the cavity,
provided that the condition
\mbox{$R_2$ $\!\ll$ $\!R_1$} is satisfied.
The transmission-line splitting
is then quite similar to the Rabi splitting mentioned above
(compare Fig.~\ref{fig2} with Figs.~3 and 4 in Ref.~\cite{Turch95}).


\section{Relation to the Interaction-free-measurement
and the Quantum Zeno effect}
\label{App-Zeno}

If one (or both) of the mirrors of the horizontal cavity is replaced by a
completely absorbing object (i.e., \mbox{$A_{M1}$ $\!=$ $\!1$} and/or
\mbox{$A_{M3}$ $\!=$ $\!1$}), then in the resonance regime
the double-cavity system changes
from an almost
perfectly transmitting device
to an almost perfectly reflecting
device.
To be more specific, while for perfect mirrors of the horizontal
cavity, in resonance the fraction of the reflected light is 
\mbox{$\approx$ $\!R_1^2/4$ $\!\ll$ $\!1$} and the fraction of the
transmitted light is \mbox{$\approx$ $\!1$ $\!-$ $\!R_1^2/4$},
for a fully absorbing horizontal cavity one gets for the transmitted
fraction \mbox{$\approx$ $\!4T_1 R_2^2/R_1^2$ $\!\ll$ $\!1$}
and the reflected fraction \mbox{$\approx$ $\!1$ $\!-$
$\!4T_1 R_2 /R_1$ $\!\approx$ $\!1$}. Thus, almost without
touching the absorber (and being lost), a photon can tell us
whether the absorber is in the  horizontal cavity or
not. The probability of such a loss is \mbox{$\approx$ $\!4T_1 R_2
(R_1$ $\!-$ $\!R_2)/R_1^2$ $\!\approx$ $\!4T_1 R_2 /R_1$} and can be,
in principle, made as small as we wish [see Fig.~(\ref{figabs})]. 
This is effect is also called
{\em interaction-free measurement\/} \cite{Vaidman}. 
It was suggested for a different
coupled cavity system (coupling by a highly reflecting, partially
transmitting mirrors) in \cite{Kwiat95} and demonstrated
experimentally (for coupling by polarization rotation) in
\cite{Kwiat97}. 

One can understand the phenomenon also in terms of the {\em quantum Zeno
effect}.  Let us consider a two-state system where a photon can be either in
the vertical or in the horizontal cavity, the coupling between the two states
being realized by BS$_2$.  The evolution starts with the photon in the vertical
cavity, while the presence of a photon in the horizontal cavity is monitored by
a detector. After a time interval $\Delta t$ (much shorter than the period of
oscillation between the two states), the two-level system evolves to a state
where with probability \mbox{$\propto$ $\!\Delta t^2$}
the photon is in the horizontal cavity.
If the detector sees no photon (which happens in most cases), then the
state is projected back into the original state. Thus, the photon is
prevented from entering the horizontal cavity by the presence of the
detector (for two-mode photonic Zeno effects in other schemes, see
\cite{Schleich}). Let us emphasize that the quantum nature
of the effect is observed on a single-photon level.     
The effect of inhibiting waves from entering the absorbing
part of the resonator can of course be explained 
classically.


\section{Intra-cavity fields}
\label{Ap-intracavity}

The field propagating from BS$_1$ to BS$_2$ has the
amplitude $a_{12}$ given by
\begin{eqnarray}
 a_{12} = ir_1 \frac{-t_1 \sqrt{1-A_{M4}}e^{i2kL_5}a+b}
{1+T_1 \sqrt{1-A_{M4}}e^{i2kL_5} {\cal B}},
\label{Ap12}
\end{eqnarray}
with $a$ and $b$ the input amplitudes in Eq. (\ref{inout}) and ${\cal B}$
given in (\ref{Bdef}). The field arriving at  BS$_1$ from  BS$_2$ 
has the amplitude  $a_{21}$ given by
\begin{eqnarray}
 a_{21}= {\cal B}a_{12} .
 \label{Ap21}
\end{eqnarray}
The field propagating from BS$_1$ to M$_4$ has the
amplitude $a_{1,M4}$ given by
\begin{eqnarray}
 a_{1,M4} = ir_1 \frac{a+t_1  {\cal B} b}
{1+T_1 \sqrt{1-A_{M4}}e^{i2kL_5} {\cal B}},
\label{Ap1M4}
\end{eqnarray}
the field arriving at  BS$_1$ from   M$_4$ has the
amplitude $a_{M4,1}$ given by
\begin{eqnarray}
 a_{M4,1}=-\sqrt{1-A_{M4}} e^{i2kL_5} a_{1,M4} .
 \label{ApM41}
\end{eqnarray}
The field propagating from BS$_2$ to M$_1$ has the
amplitude $a_{2,M1}$ given by
\begin{eqnarray}
 a_{2,M1} = -ir_2 t_2
\left( \sqrt{1-A_{M3}}e^{i2kL_2}
\right. \nonumber \\
\left. +\sqrt{1-A_{M2}}e^{i2kL_4}  
\right) {\cal B}_5^{-1} e^{ikL_1} a_{12}, 
\label{Ap2M1}
\end{eqnarray}
where
\begin{eqnarray}
{\cal B}_5 = 1- \sqrt{1-A_{M1}}e^{i2kL_3} \left(  T_2 \sqrt{1-A_{M3}}
e^{i2kL_2} 
\right. \nonumber \\
\left. - R_2 \sqrt{1-A_{M2}} e^{i2kL_4} \right) ;
\label{AB5}
\end{eqnarray}
the field arriving at  BS$_2$ from   M$_1$ has the
amplitude $a_{M1,2}$ given by
\begin{eqnarray}
 a_{M1,2}=-\sqrt{1-A_{M1}} e^{i2kL_3} a_{2,M1} .
 \label{ApM12}
\end{eqnarray}
The field propagating from BS$_2$ to M$_2$ has the
amplitude $a_{2,M2}$ given by
\begin{eqnarray}
 a_{2,M2} =  t_2
\left( 1- \sqrt{(1-A_{M1})(1-A_{M3})}
\right. \nonumber \\
\left. \times 
(1-A_{2})e^{i2k(L_2+L_3)}  
\right) 
{\cal B}_5^{-1} e^{ikL_1} a_{12}, 
\label{Ap2M2}
\end{eqnarray}
the field arriving at  BS$_2$ from M$_2$ has the
amplitude $a_{M2,2}$ given by
\begin{eqnarray}
 a_{M2,2}=-\sqrt{1-A_{M2}} e^{i2kL_4} a_{2,M2} .
 \label{ApM22}
\end{eqnarray}
The field propagating from BS$_2$ to M$_3$ has the
amplitude $a_{2,M3}$ given by
\begin{eqnarray}
 a_{2,M3} =  -ir_2
\left( 1+ \sqrt{(1-A_{M1})(1-A_{M2})}
\right. \nonumber \\
\left.
\times (1-A_{2})e^{i2k(L_3+L_4)}  
\right) {\cal B}_5^{-1} e^{ikL_1} a_{12},
\label{Ap2M3}
\end{eqnarray} 
the field arriving at  BS$_2$ from   M$_3$ has the
amplitude $a_{M3,2}$ given by
\begin{eqnarray}
 a_{M3,2}=-\sqrt{1-A_{M3}} e^{i2kL_2} a_{2,M3} .
 \label{ApM32}
\end{eqnarray}




\begin{references}

\bibitem{Schm96}
H. Schmidt and A. Imamo\u{g}lu,
Opt. Lett. {\bf 21,} 1936 (1996).

\bibitem{Schm98}
H. Schmidt and A. Imamo\u{g}lu,
Opt. Lett. {\bf 23,} 1007 (1998).

\bibitem{Harris90}
S.E. Harris, J.E. Field, and A. Imamo\u{g}lu,
Phys. Rev. Lett. {\bf 64,} 1107 (1990).

\bibitem{Harris97}
S.E. Harris,
Phys. Today {\bf 50(7),} 36 (1997).

\bibitem{Maran98}
J.P. Marangos,
J. Mod. Optics {\bf 45,} 471 (1998).

\bibitem{Imam97}
A. Imamo\u{g}lu, H. Schmidt, G. Woods, and M. Deutsch,
Phys. Rev. Lett. {\bf 79,} 1467 (1997).

\bibitem{Harris99}
S.E. Harris and L.V. Hau
Phys. Rev. Lett. {\bf 82,} 4611 (1999).

\bibitem{Lukin00}
M.D. Lukin and A. Imamo\u{g}lu, 
Phys. Rev. Lett. {\bf 84,} 1419 (2000).

\bibitem{Lukin98}
M.D. Lukin, M. Fleischauer, M.O. Scully, and V.L. Velichansky, 
Opt. Lett. {\bf 23,} 295 (1998).


\bibitem{Gran98}
Ph. Grangier, D.F. Walls, and K.M. Gheri,
Phys. Rev. Lett. {\bf 81,} 2833 (1998).

\bibitem{Imam98E}
A. Imamo\u{g}lu, H. Schmidt, G. Woods, and M. Deutsch,
Phys. Rev. Lett. {\bf 81,} 2836 (1997).

\bibitem{Gheri99}
K.M. Gheri, W. Alge, and Ph. Grangier, 
Phys. Rev. A {\bf 60,} R2673 (1999).


\bibitem{Kuhn}
Axel Kuhn, private communication.


\bibitem{Hau99}
L.V. Hau, S.E. Harris, Z. Dutton, and C.H. Berhoozi,
Nature {\bf 397,} 594 (1999).

\bibitem{Mara99}
J. Marangos,
Nature {\bf 397,} 559 (1999).

\bibitem{Stefan}
L. Kn\"{o}ll, S. Scheel, E. Schmidt, D.-G. Welsch, and A.V. Chizhov,
Phys. Rev. A {\bf 59,} 4716 (1999).

\bibitem{EITsolid}
B.S. Ham, P.R. Hemmer, M.S. Shahriar,
Opt. Commun. {\bf 144,} 227 (1997);
B.S. Ham, S.M. Shahriar, and P.R. Hemmer,
JOSA B {\bf 16,} 801 (1999);
A.V. Turukhin, J.A. Musser, V.S. Sudarshanam, M.S. Shahriar,
and P.R. Hemmer,
e-print quant-ph/0010009.

\bibitem{Fortphys}
Fortschritte der Physik, Special Issue on Experimental Proposlals for Quantum
Computation, Eds. S. Braunstein and Hoi-Kwong Lo, vol. {\bf 48,} pp. 765-1140 
(2000).

\bibitem{ParisVitali}
M.G.A. Paris, M. Plenio, D. Jonathan, S. Bose, and G.M. D'Ariano,
Phys. Lett. A {\bf 273,}  153  (2000);
D. Vitali, M. Fortunato, and P. Tombesi,
Phys. Rev. Lett. {\bf 85,} 445 (2000).

\bibitem{Kash99}
M.M. Kash, V.A. Sautenkov, A.S. Zibrov, L. Hollberg, G.R. Welch, M.D. Lukin, Yu.
Rostovtsev, E.S. Fry, and M.O. Scully,
Phys. Rev. Lett. {\bf 82,} 5229 (1999).

\bibitem{Teich}
B.A.E. Saleh and M.C. Teich, {\em Fundamentals of Photonics.\/} (Wiley, New
York, 1991), Chapt. 3.

\bibitem{Lukin-private}
M.D. Lukin, private communication.

\bibitem{Ulf}
U. Leonhardt and P. Piwnicki,
Phys. Rev. Lett. {\bf 84,} 822 (2000).

\bibitem{Turch95}
Q.A. Turchette, R.J. Thompson, and H.J. Kimble,
Appl. Phys. B {\bf 60,} S1 (1995).

\bibitem{Vaidman}
A.C. Elitzur and L. Vaidman, Found. Phys. {\bf 23,} 987 (1993).

\bibitem{Kwiat95} 
P. Kwiat, H. Weinfurter, T. Herzog, A. Zeilinger, and M.A. Kasevich,
Phys. Rev. Lett. {\bf 74,} 4763 (1995).


\bibitem{Kwiat97}
P.G. Kwiat,
Phys. Scr. {\bf T76,} 115 (1998);
P.G. Kwiat, A.G. White, J.R. Mitchell, O. Nairz, G. Weihs, H. Weinfurter, and
A. Zeilinger,
Phys. Rev. Lett. {\bf 83,} 4725 (1999).

\bibitem{Schleich}
W.P. Schleich, P.J. Bardroff, and H. Walther, {\em Zeno effect, interaction free
measurement, and causality with two coupled field modes,\/} preprint,
Verhandlungen der Deutschen
Physikalischen Gesellschaft, Bonn 2000; A.G. Kofman, G. Kurizki, and T.
Opatrn\'{y}, Phys. Rev. A {\bf 63,} 042108 (2001).


\end{references}
\end{document}